# Mechanical oscillations in lasing microspheres

A.Toncelli[1], N. E. Capuj[2], B. Garrido[3], M. Sledzinska[4], C. M. Sotomayor-Torres[4,5], A. Tredicucci[1,6], D. Navarro-Urrios[a),4]

[1]*NEST, CNR Istituto Nanoscienze and Dipartimento di Fisica, Università di Pisa, Largo Pontecorvo 3, 56127 Pisa, Italy.*

[2]*Depto. Física, Universidad de la Laguna, La Laguna, Spain.*

[3]*MIND-IN2UB, Departament d'Electrònica, Facultat de Física, Universitat de Barcelona, Martí i Franquès 1, 08028 Barcelona, Spain*

[4]*Catalan Institute of Nanoscience and Nanotechnology (ICN2), CSIC and The Barcelona Institute of Science and Technology, Campus UAB, Bellaterra, 08193 Barcelona, Spain.*

[5]*Catalan Institute for Research and Advances Studies ICREA, Barcelona, Spain.*

[6]*NEST, CNR Istituto Nanoscienze and Scuola Normale Superiore, Piazza San Silvestro 12, 56127 Pisa, Italy.*

We investigate the feasibility of activating coherent mechanical oscillations in lasing microspheres by modulating the laser emission at a mechanical eigenfrequency. To this aim, 1.5%$Nd^{3+}$:Barium-Titanium-Silicate microspheres with diameters around 50 μm were used as high quality factor ($Q>10^6$) whispering gallery mode lasing cavities. We have implemented a pump-and-probe technique in which the pump laser used to excite the $Nd^{3+}$ ions is focused on a single microsphere with a microscope objective and a probe laser excites a specific optical mode with the evanescent field of a tapered fibre. The studied microspheres show monomode and multi-mode lasing action, which can be modulated in the best case up to 10 MHz. We have optically transduced thermally-activated mechanical eigenmodes appearing in the 50-70 MHz range, the frequency of which decreases with increasing the size of the microspheres. In a pump-and-probe configuration we observed modulation of the probe signal up to the maximum pump modulation frequency of our experimental setup, i.e., 20 MHz. This modulation decreases with frequency and is unrelated to lasing emission, pump scattering or thermal effects. We associate this effect to free-carrier-dispersion induced by multiphoton pump light absorption. On the other hand, we conclude that, in our current experimental conditions, it was not possible to resonantly excite the mechanical modes. Finally, we discuss on how to overcome these limitations by increasing the modulation frequency of the lasing emission and decreasing the frequency of the mechanical eigenmodes displaying a strong degree of optomechanical coupling.

[a)] Corresponding author: e-mail daniel.navarro@icn2.cat, Phone: + 34 93 737 2641

## I. INTRODUCTION

Optical microcavities have been proposed for applications in different sensing schemes [1]-[3], all-optical switches [4] and micro-mechanical oscillator systems [5], among others.

In such microcavity structures, light is confined to a small volume by means of total internal reflection. Different shapes, such as microdisks, microrings, microtoroids and microspheres give rise to whispering gallery mode (WGM) resonances that localise the radiation beneath the surface of the microresonator [6]. If the microresonator is composed of a light emitting medium, the luminescent spectrum is modulated by the WGM spectrum, and the spontaneous emission rate might even be dramatically enhanced [7].

Microspheres (μS) are versatile objects that can be made of many materials including polymers, crystalline and glassy compounds. Moreover, quality factors (Q) up to $8x10^9$ can be achieved in this geometry [8]. Such ultra-high Qs lead to very long light storage times in the microcavity, which enables lasing action at low pump power thresholds if an active medium providing optical gain is embedded in the μS [9]. Lasing cavities can surpass the high sensitivity of passive WGM cavities and achieve extremely low detection limits thanks to their narrow linewidths [10]

In this work, we use optically active Nd-doped glass microspheres pumped above the lasing threshold in a pump-and-probe setup to explore the possibility of activating coherent mechanical oscillations in WGM microresonators. The main idea is to exploit a quite unexplored concept in optomechanics, which is the use of the lasing cavities to achieve very high electromagnetic field intensities to enhance radiation pressure forces. Moreover, this strategy allows to overcome the stringent requirements that passive resonators must satisfy in terms of quality factor to modify the intrinsic mechanical relaxation rates. Indeed, laser linewidths can be many orders of magnitude narrower than those associated to the passive cavity modes. Lasing action combined with coherent mechanical oscillation could find interesting applications, for instance, as dual sensor of two independent material properties [11].

In our experiment, the pump light is used to excite the $Nd^{3+}$ ions and the probe beam monitors an optical resonance in the spectral region around 1.5 μm. The radiofrequency (RF) spectrum of the optical signal from the μS has been analysed in pump-on/probe-off, pump-off/probe-on, pump-on/probe-on configurations.

## II. EXPERIMENTAL

### A. Microsphere preparation

The Nd-doped Barium-Titanium-Silicate μS were fabricated from a glass with the composition of 40%BaO–20%$TiO_2$–40%$SiO_2$ and doped with 1.5% $Nd_2O_3$ (in molar ratio). The glass was reduced to dust in a mortar and subsequently heated up to its fusion temperature, which is around 900 °C. Most of the splinters melt and, when the temperature decreases, solidify in spherical shapes of several micrometer radii, thus becoming high-Q WGM lasing cavities. Details of the experimental methods can be found elsewhere [7]. The μS were, then, placed on a microscope glass.

### B. Pump-and-probe experimental setup

The pump-and-probe setup used in this work is shown in Figure 1. A continuous wave 808 nm pump laser (Dragon Lasers M-series 2 Watts) was focused with a microscope objective (Mitutoyo 50X) on a single μS in a top pumping configuration. The resulting focused beam had a spot area of about $1 \times 10^{-6}$ $cm^2$, measured by comparing it with a structure of known size. The pump laser was modulated with an Electro-Optic-Modulator (EOM) tunable in a frequency range between 0.5 and 20 MHz. The relative position of the single μS with respect to the pump beam could be adjusted with a xyz micro-precision stage. Under external excitation, $Nd^{3+}$ ions are promoted from the ground state ($^4I_{9/2}$) up to the $^4F_{5/2}$ and $^2H_{9/2}$ levels, from where a rapid non-radiative thermalization to the $^4F_{3/2}$ transition takes place. Laser action at around 1.07 μm occurs from the $^4F_{3/2} \rightarrow ^4I_{11/2}$ transition provided that internal gain compensates passive losses. $Nd^{3+}$ effective dynamics can be modelled by a four-level rate equation system.

A tapered fibre with a diameter of about 1μm was brought into contact with the μS surface. The fibre provided the means to both couple a probe laser light to a WGM and to collect the laser emission from the μS. The probe light was provided by a tunable laser source at around 1.5 μm (Tunics T100S-HP), far from any optical absorption of the active centres. The probe laser was coupled to the fibre input. Both the transmitted laser signals and the photoluminescence (PL) emission could be detected at the fibre output.

To check for the presence of a radiofrequency (RF) modulation of the transmitted signal an InGaAs fast photoreceiver with a bandwidth of 12 GHz was used. The RF voltage was connected to the 50 Ohm input impedance of an electrical spectrum analyzer (ESA) (Anritsu-MS2830A) with a bandwidth of 13.5 GHz. The whole setup operated at atmospheric conditions of temperature and pressure.

In order to perform the optical spectral analysis, an Optical Spectrum Analyzer (OSA) (Instrument Systems Spectro 320) spanning the visible and near-IR region was placed after the fibre output.

## III. RESULTS

We characterized three different μS with different diameters labelled as S1 (d~40 μm), S2 (d~50 μm), S3 (d~55 μm), respectively.

### A. CW Photoluminescence measurements: Pump-on/probe-off configuration

In the configuration with the probe laser off, light emitted by the μS is directly out-coupled to the tapered fibre and analysed with the OSA. The three μS under analysis achieved the lasing regime. In particular, we made a study of the PL emission as a function of the pump photon flux ($\Phi_p$) in S1 (Fig. 2), which was the one showing the lowest threshold for lasing (inset to Fig. 2) and multimode lasing behaviour. It is possible to observe the scattering of the pump laser at around 808 nm and the PL emission associated to the $^4F_{3/2} \rightarrow {}^4I_{9/2}$ and the $^4F_{3/2} \rightarrow {}^4I_{11/2}$ transitions, the latter being the one showing laser action.

### B. CW transmission measurements: Pump-off/probe-on configuration

A typical transmission spectrum of the probe light in a wide spectral range when using a low probe power (about 80 μW) is shown in Fig. 3a. Dips in the transmitted signal are visible every time the laser is resonant with a WGM. Interestingly, the dips are composed by a complex structure of modes. Those are associated to azimuthal WGMs whose energy degeneracy has been broken by eccentricity splitting. In many of the observed modes (see Fig. 3b), the optical Qs are limited by the resolution of the experimental setup, thus exceeding $10^6$.

It is worth noting that, when the pump laser is switched on, part of the absorbed light leads to homogeneous heating effects and induces thermo-optic (TO) nonlinearities, i.e., a spectral red-shift of the transmission spectrum. At the highest powers, the spectral shift can be as high as 2 nm. When the probe power is increased, the excited WGM redshifts as a consequence of TO nonlinearities as well. However, in this case the heating source is localized in the WGM volume.

### C. RF measurements: Pump-off/probe-on configuration

In the absence of the pump laser, thermally driven motion of the low frequency mechanical modes supported by the μS can be seen by processing the transmitted light with the ESA. Mechanical modes with a non-negligible optomechanical coupling rate appear as narrow peaks in the RF spectrum, as reported in Fig. 4a. As expected, the mechanical mode frequency decreases with increasing the μS size, the mechanical quality factor being of the order of $10^2$ in all cases. The latter values are mainly limited by the experimental conditions: room-temperature (thermo-elastic damping) and atmospheric pressure. Not all the WGMs show enough optomechanical coupling to transduce mechanical eigenmodes efficiently enough to overcome the noise level of our RF measurements. For each characterized μS, we have excited the WGM that enables the highest transduced RF signal. We are able to transduce modes whose spectral position is in agreement with what expected for the fundamental breathing modes of the μS. Interestingly, in some cases there seems to be more than one mode, which are non-degenerate breathing-like modes arised as a consequence of the μS eccentricity.

### D. RF measurements: Pump-and-probe configuration

In the following, we report on the RF spectra displayed by S1 under: i) the pump-off/probe-on, ii) the pump-on/probe-off and iii) the pump-on/probe-on experimental conditions. All experiments have been performed at the maximum available pump flux (about $2x10^{24}$ phot $cm^{-2}$ $s^{-1}$) and probe power (2 mW). Fig. 4 b-d show an example of these results at a modulation frequency of 2.4 MHz.

The pump-off/probe-on (Fig. 4b) configuration corresponds to that discussed in section C and the black curve of Fig. 4a, where the RF spectrum displays the mechanical modes that have a sufficient degree of optomechanical coupling rate and are thermally activated. In this case, the set of transduced mechanical modes appear slightly above 70 MHz, i.e., 3.5 times higher than the maximum modulation frequency able to be generated by the EOM. Interestingly, we are able to transduce a set of mechanical modes that are very close in frequency instead of a single mechanical breathing mode typical from a perfect μS. This is probably associated to the splitting of the latter mode into different breathing-like modes of a geometrically-perturbed μS.

The pump-on/probe-off (Fig. 4c) configuration gives rise to RF peaks at the modulation frequency of the pump beam and its harmonics. In the case reported the main peak appears at 2.4 MHz and three harmonics are clearly visible. As it will be justified later on in this section, these RF peaks are associated to modulation of the laser emission at 1.07 μm, meaning that the μS laser emission is being switched on and off periodically at the EOM frequency. It is worth noting that only the first harmonic lies above the noise floor for modulation frequencies above 4 MHz (not shown).

The pump-on/probe-on situation (Fig. 4d) also shows RF peaks at the modulation frequency of the pump beam and its harmonics, but the intensity of the peaks is much higher than in the pump-on/probe-off configuration. This means that the modulated pump induces a clear modulation of the probe in addition to the modulation of the μS laser emission discussed above. The modulation on the probe is also periodic, but deviates from a sinusoidal shape in a way that there appear multiple harmonics of a main peak at the frequency of EOM.

The probe modulation mechanism when the pump is on can be understood in the following terms. The presence of the pump induces either a refractive index and/or a geometrical change in the μS, leading to an optical shift of the WGM that is being probed. This in turn modifies the transmitted intensity of the probe laser, which is modulated by the lorentzian shape of the optical resonance. Even under the effect of the pump, the probe beam is still able to transduce the thermally activated mechanical modes. The latter signal is slightly lower than in the case of Fig. 4b because the modulation induced by the pump deviates the optical resonance from the optimum condition for transduction. The spectral shape of the mechanical signal is also slightly different because the probe laser can be also coupled to other WGM placed close by, which may show slightly different optomechanical coupling rates with the bunch of breathing-like modes.

Finally, we performed a study of the RF spectra as a function of the modulation frequency. In Fig. 4e we plot the main peak intensity for the pump-on/probe-off con-figuration (black dots) and for the pump-on/probe-on con-figuration (red dots). The peak intensity in the pump-on/probe-off decreases with frequency and completely quenches at about 10 MHz, which implies that it is associated to laser emission and not to pump scattering, whose frequency response should be flat. Indeed, the scattered contribution, which is observed in the OSA (see Fig. 2), is hidden below the noise of the ESA. The maximum achievable modulation frequency of the laser emission is directly related to the rate at which the higher $Nd^{3+}$ energy levels are populated and depopulated. Therefore, in order to further increase these rates, a valid strategy would be to increase the pump photon flux, either by focusing to a smaller spot or by increasing the laser power. With the current experimental setup, we are working at increasing the maximum photon flux available. It is also worth noting that the spot size could be decreased only to the size of the single μS.

In the modulation frequency range our system could span, we have not observed any clear enhancement of the signal associated to the mechanical modes. Indeed, this is reasonable since the frequency of the mechanical modes observable through optomechanical coupling is much higher than the maximum frequency at which the lasing emission can be modulated, preventing any resonant actuation by using the main or higher harmonics of the radiation pressure force.

In a pump-on/probe-on configuration, the main RF peak intensity also decreases with frequency, but it is always more intense than in a pump-on/probe-off configuration and at low frequency it shows up to six clear harmonics (only four in the case of Fig. 4d). The RF signal decrease is also much smoother, especially at high frequencies. As a result, it is still clearly visible up to the maximum modulation frequency available by our system (20 MHz). This strong modulation of the probe signal cannot be related to lasing or to pump scattering. We also rule out thermal effects since their associated dynamics are usually much slower (in the KHz range) [12]. Therefore, we associate this effect to another non-linear mechanism that modulates the effective refractive index of the WGM. This is probably free-carrier dispersion in response to a multiphoton absorption process. Future work will assess whether this modulation mechanism can be brought to the range in which the mechanical modes appear and induce mechanical amplification or lasing.

Regarding the other two studied μS, both S2 and S3 showed cut-off frequencies for the lasing emission around 1MHz. On the other hand, the pump-and-probe signal was in both cases much weaker than that of S1, but remained visible up to 20 MHz as well.

## IV. FINITE ELEMENT SIMULATIONS

Single-particle optomechanical coupling rates ($g_{o,OM}$) between optical WGM and mechanical modes are estimated by taking into account the moving interface (MI) effects [13-15], which come from the dielectric permittivity variation at the boundaries associated with the deformation.

The calculation of the MI coupling coefficient $g_{MI}$ is performed using the integral given by Johnson et al. [13] evaluated in spherical coordinates;

$$g_{MI} = -\frac{\pi\lambda_r}{c}\frac{\oint (\boldsymbol{Q}\cdot\hat{\mathbf{n}})\left(\Delta\varepsilon \boldsymbol{E}_{\parallel}^2 - \Delta\varepsilon^{-1}\boldsymbol{D}_{\perp}^2\right)dS}{\int \boldsymbol{E}\cdot\boldsymbol{D}dV}\sqrt{\hbar / 2m_{eff}\Omega_m} \tag{1}$$

where **Q** is the normalized displacement (max{|**Q**|}=1), $\hat{\boldsymbol{n}}$ is the normal at the boundary (pointing outward), **E** is the electric field and **D** the electric displacement field. $\varepsilon$ is the dielectric permittivity, $\Delta\varepsilon=\varepsilon_{silica}-\varepsilon_{air}$ , $\Delta\varepsilon^{-1}=\varepsilon^{-1}_{silica}-\varepsilon^{-1}_{air}$. $\lambda_r$ is the optical resonance wavelength, c is the speed of light in vacuum, $\hbar$ is the reduced Planck constant, $m_{eff}$ is the effective mass of the mechanical mode and $\Omega_m$ is the mechanical mode eigenfrequency, so that $\sqrt{\hbar / 2m_{eff}\Omega_m}$ is the zero-point motion of the resonator. It is worth noting that, in our model, we have imposed a rotational symmetry around the vertical axis of the μS.

Fig. 5 shows the calculated coupling rates for a μS with a diameter of 55 μm. The highest calculated value corresponds to the interaction with the fundamental breathing mode of the μS, $g_{MI}/2\pi$ being 38 Hz. This particular mechanical mode appears slightly below 70 MHz for a μS diameter of 55 μm. The slight disagreement with the experimental value is probably associated with the elastic constant values imposed in our modelling, which have been taken as if the glass matrix was silica. The rather low value of $g_{MI}/2\pi$ is a consequence of the large volume of the mechanical mode in comparison to the optical WGM under consideration.

## V. CONCLUSIONS

We thoroughly optical characterized lasing glass microspheres of different diameters using a pump-and-probe experimental setup, aiming to explore the feasibility of resonantly exciting a mechanical mode by modulating the laser emission. We demonstrated that the lasing mechanism can be modulated up to about 10 MHz in the best studied case. However, the mechanical modes displaying strong optomechanical coupling, i.e., susceptible to be efficiently driven by optical forces, are measured between 50 and 70 MHz in all studied cases, which prevent achieving resonant excitation. Furthermore, we have unveiled another non-linear mechanism that modulates the probe signal, the time response of which is high enough to follow the modulation frequency until the end of the available range of our experimental setup (20 MHz). We associate this mechanism to a free-carrier dispersion mechanism generated by multiphoton absorption of the optical pump.

In order to succeed on activating mechanical coherent oscillations in glass microspheres possible strategies include either the acceleration of the dynamics of the lasing emission or a decrease of the eigenfrequency of the mechanical modes. Work is in progress to increase the pumping laser power and/or replace the Barium-Titanium-Silicate glass material by another suitable glass with lower elastic constants.

## VI. FIGURES

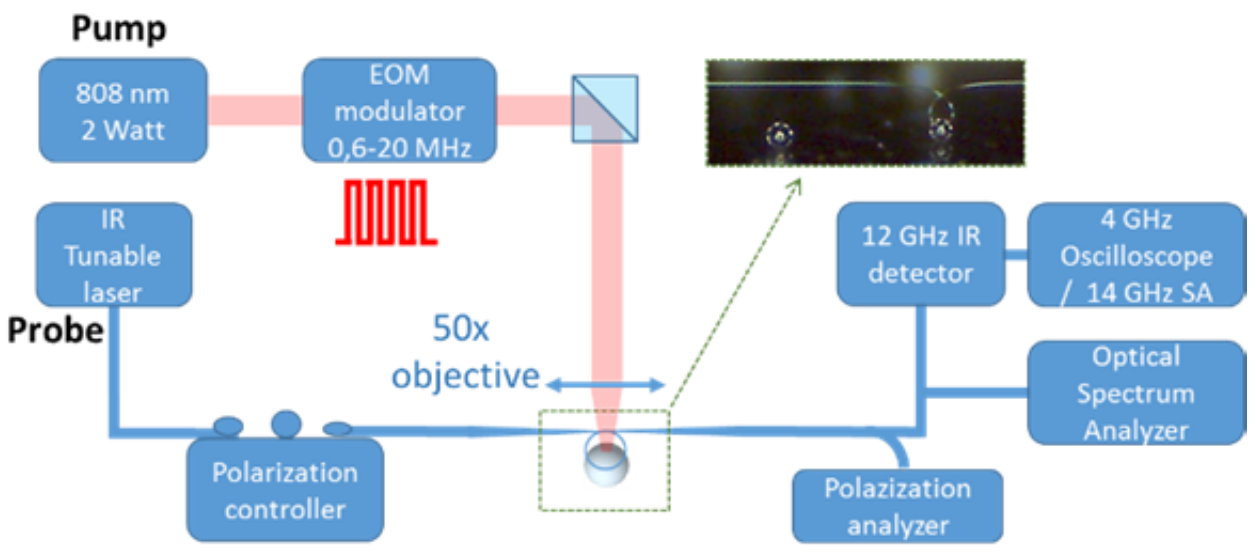

FIG. 1. Experimental pump-and-probe setup. The top right image corresponds to a side-view of the tapered fibre brought into contact with a microsphere.

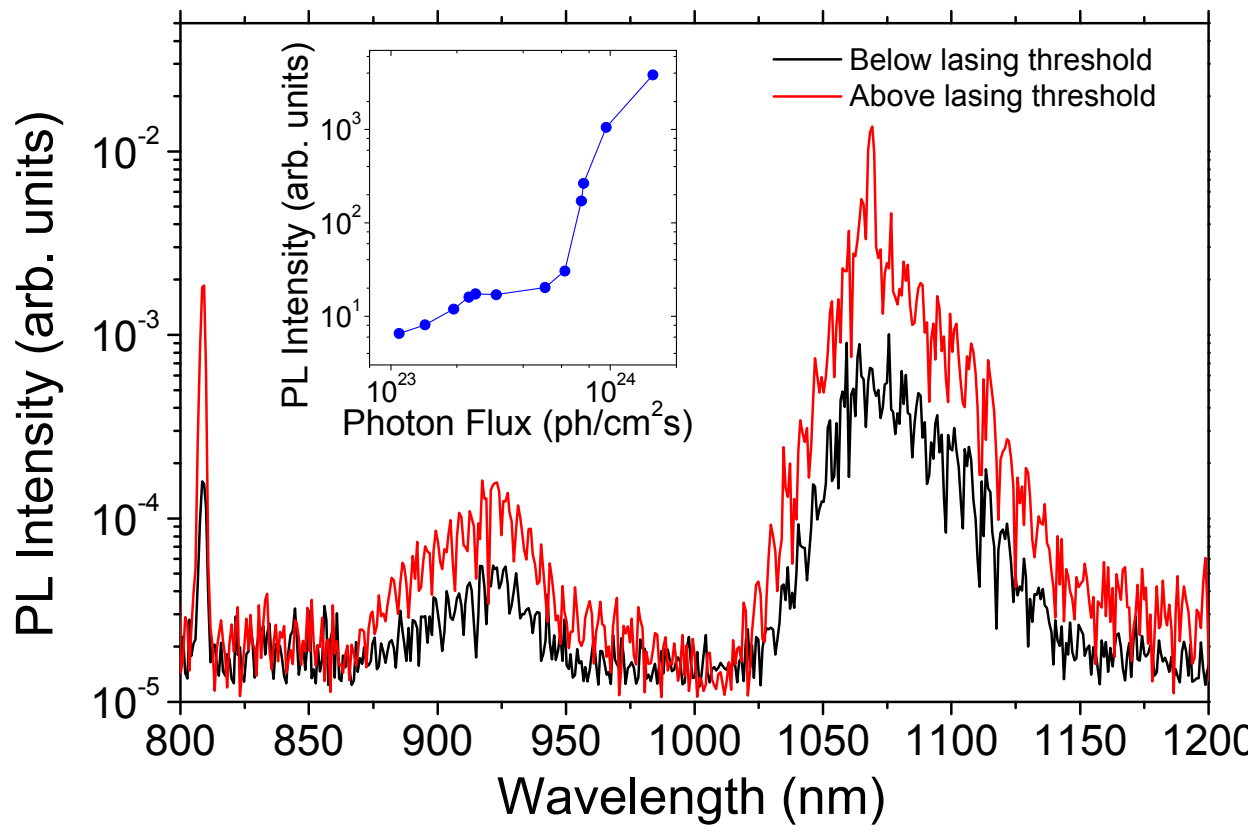

FIG. 2. PL spectra of S1 below and above the lasing threshold (black and red curves respectively). The vertical axis is in log scale. Inset. Peak intensity of a lasing mode as a function of the pump photon flux.

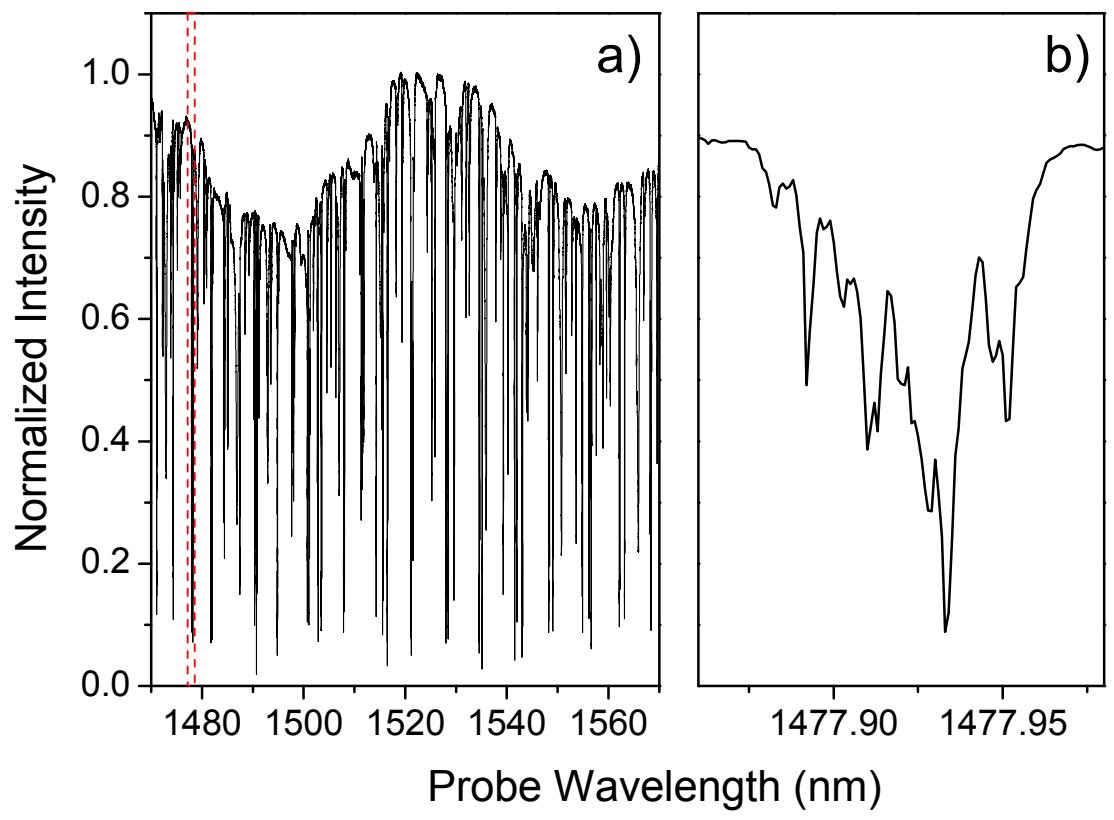

FIG. 3. a) Typical transmission spectrum of one of the measured spheres (S2 in this case). b) Zoom of one of the bunch of equivalent modes appearing in the transmission spectrum region within the dashed line box in a).

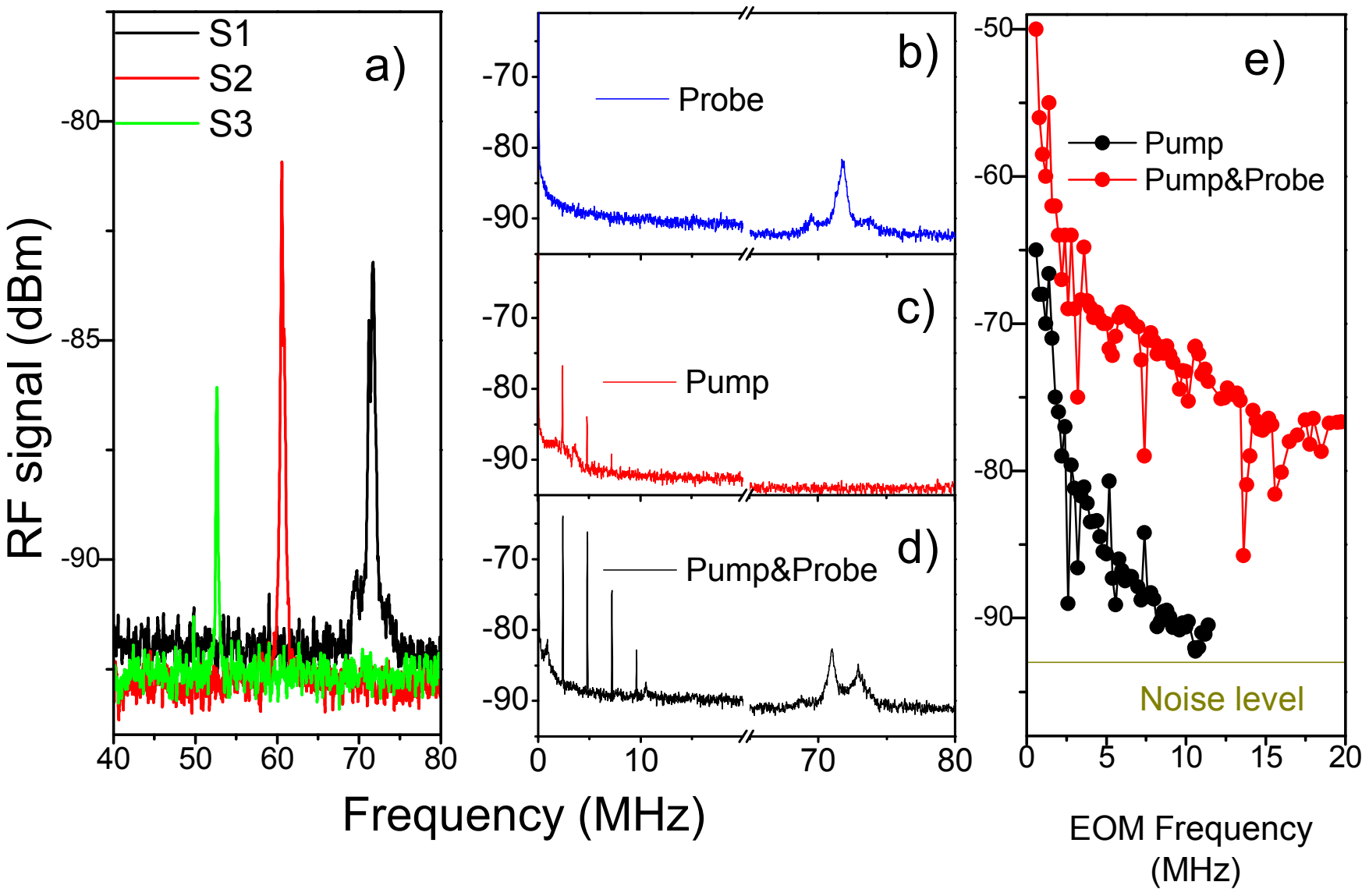

FIG. 4. a) Transduced mechanical modes of the studied microspheres. b-d) RF spectra obtained with 2.4 MHz modulation frequency in the three possible configurations: pump-off/probe-on (panel b), pump-on/probe-off (panel c) and pump-on/probe-on (panel d). e) RF main peak intensity as a function of the modulation frequency in the pump-on/probe-off (black curve) and in the pump-on/probe-on configuration.

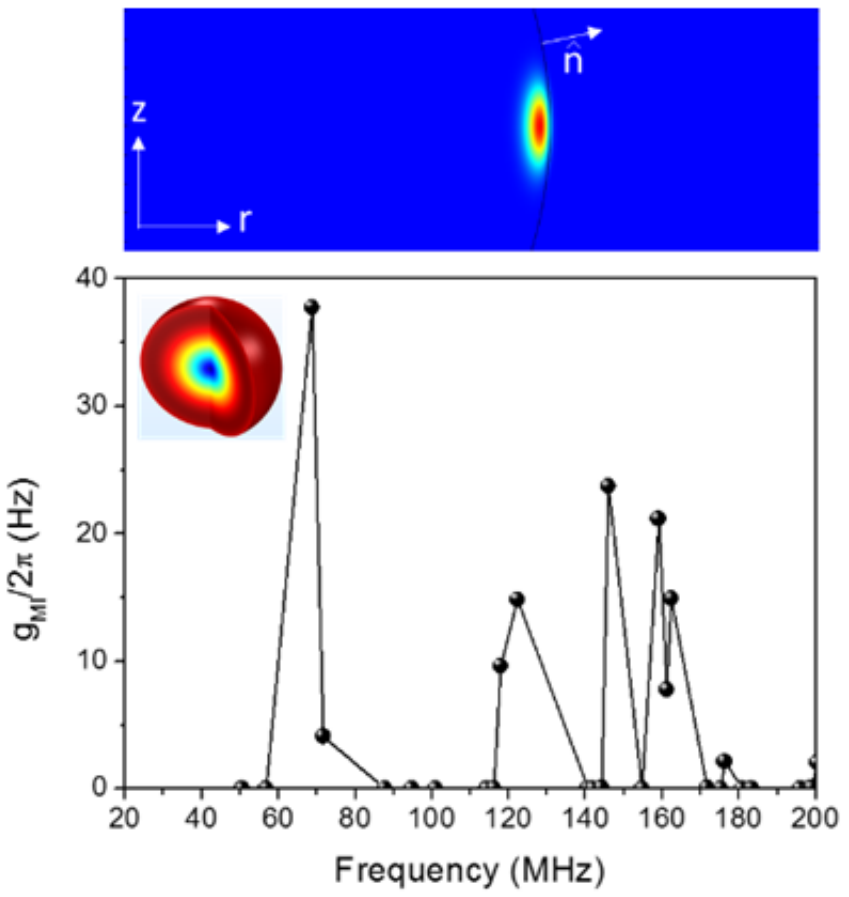

FIG. 5. a) Moving boundary optomechanical coupling rates for a sphere of a diameter of 55 μm as calculated from FEM simulations.

## ACKNOWLEDGEMENTS

The authors acknowledge M. F. Colombano for fruitful discussions. This work was supported by the European Comission project PHENOMEN (H2020-EU-713450), the Spanish Severo Ochoa Excellence program and the MINECO project PHENTOM (FIS2015-70862-P). Partial financial support by the EU through the ERC Advanced Grant SouLMan (GA 321122) is also gratefully acknowledged. DNU gratefully acknowledges the support of a Ramón y Cajal postdoctoral fellowship (RYC-2014-15392).

## REFERENCES

[1] F. Vollmer and S. Arnold, Nature Methods 5, 591 (2008)

[2] N.M. Hanumegowda, C.J. Stica, B.C. Patel, I. White and X.D. Fan, Appl. Phys. Lett. 87, 201107 (2005)

[3] C.H. Dong, L. He, Y.F. Xiao, V.R. Gaddam, S.K. Ozdemir, Z.F. Han, G.C. Guo and L. Yang, Appl. Phys. Lett. 94, 231119 (2009)

[4] S. Roy, M. Prasad, J. Topolancik and F. Vollmer, J. Appl. Phys. 107, 053115 (2010)

[5] G. Bahl, J. Zehnpfennig, M. Tomes and T. Car-mon, Nature Commun. 2, 403 (2011)

[6] A. N. Oraevsky, "Whispering gallery waves," Quantum Electron. 32, 377 (2002).

[7] L.L. Martın, D. Navarro-Urrios, F. Ferrarese-Lupi, C. Perez-Rodrıguez, I.R. Martın, J. Montserrat, C. Dominguez, B. Garrido and N. Capuj, Laser Phys. 23, 075801 (2013)

[8] K.J. Vahala, Nature 424, 839 (2003)

[9] Murugan, G., Zervas, M., Panitchob, Y. & Wilkinson, J. Integrated Nd-doped borosilicate glass microsphere laser. Opt. Lett. 36, 73 (2011).

[10] Yang, J. & Guo, L. J. Optical sensors based on active microcavities. IEEE J. Sel. Top. Quantum Electron. 12, 143 (2006).

[11] Lucklum, R.; Zubtsov, M.; Oseev, A., Phoxonic crystals—a new platform for chemical and bio-chemical sensors. Analytical and Bioanalytical Chemistry 405, 6497 (2013).

[12] Ramírez, J. M.; Navarro-Urrios, D.; Capuj, N. E.; Berencén, Y.; Pitanti, A.; Garrido, B.; Tredicucci, A., Far-field characterization of the thermal dynamics in lasing microspheres. Scientific Reports 5, 14452, (2015).

[13] S. G. Johnson et al., Phys. Rev. E 65, 066611(2002).

[14] J. Chan et al. Appl. Phys. Lett. 101, 081115, (2012).

[15] Y. Pennec et al., Nanophotonics 3(6), 413-440 (2014).

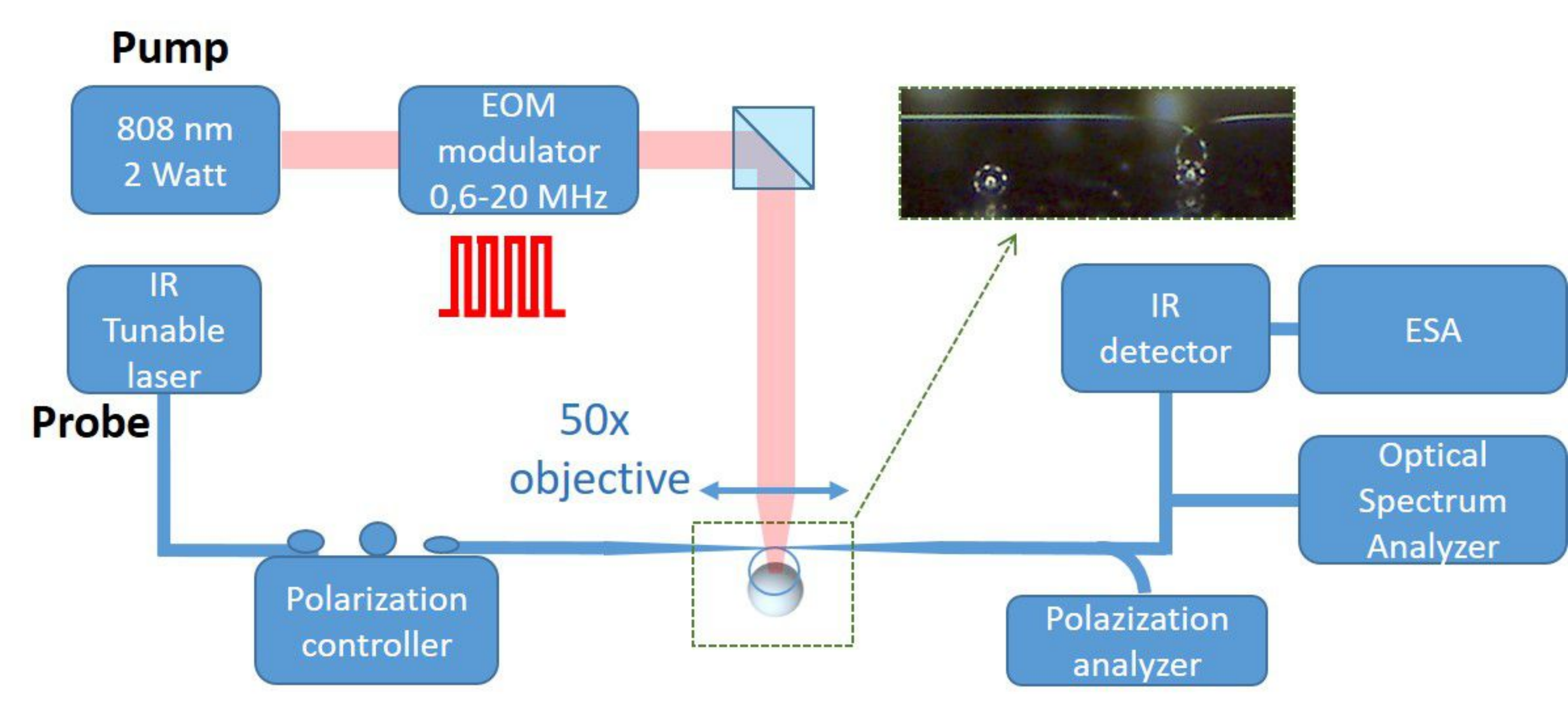

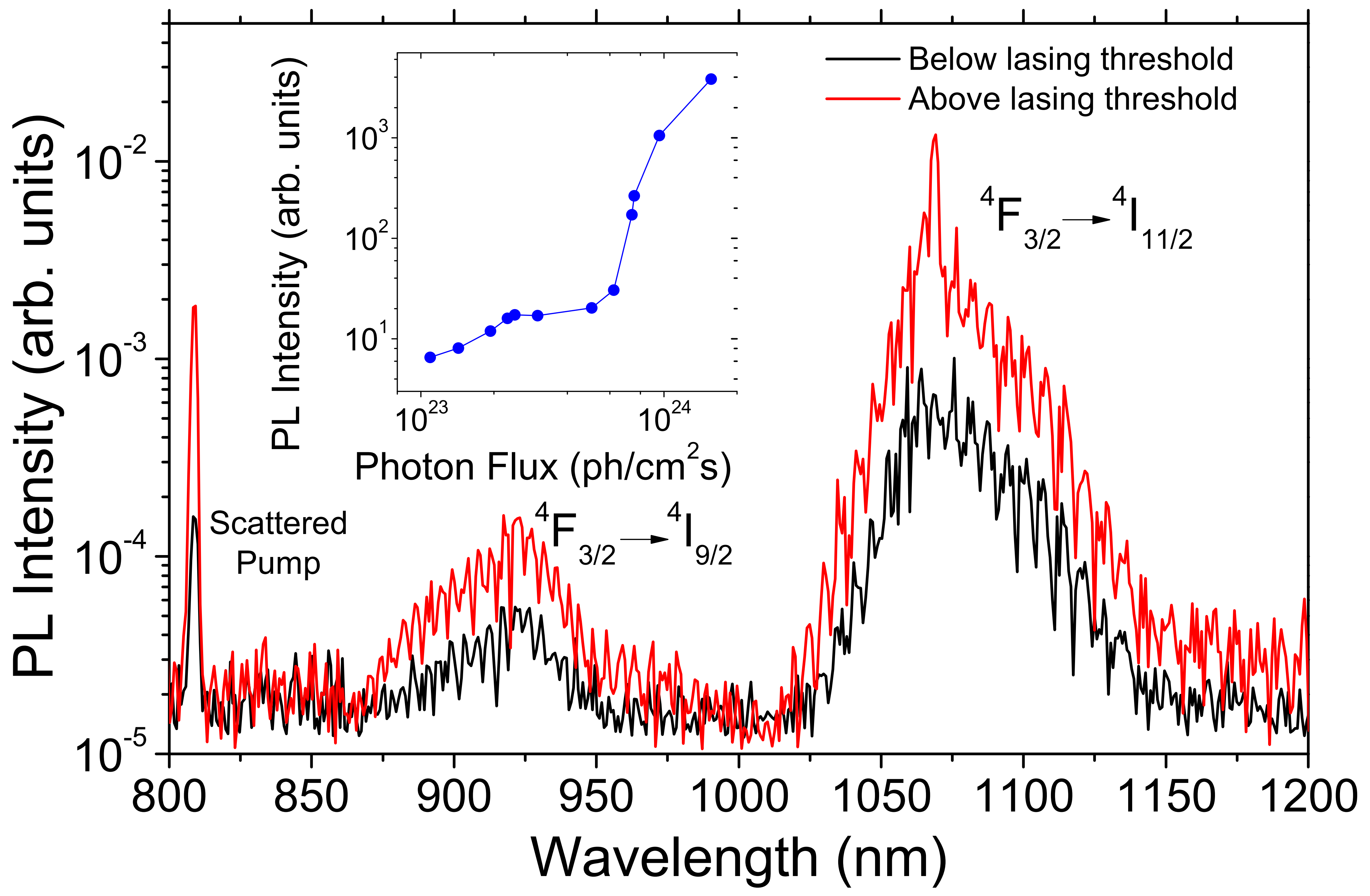

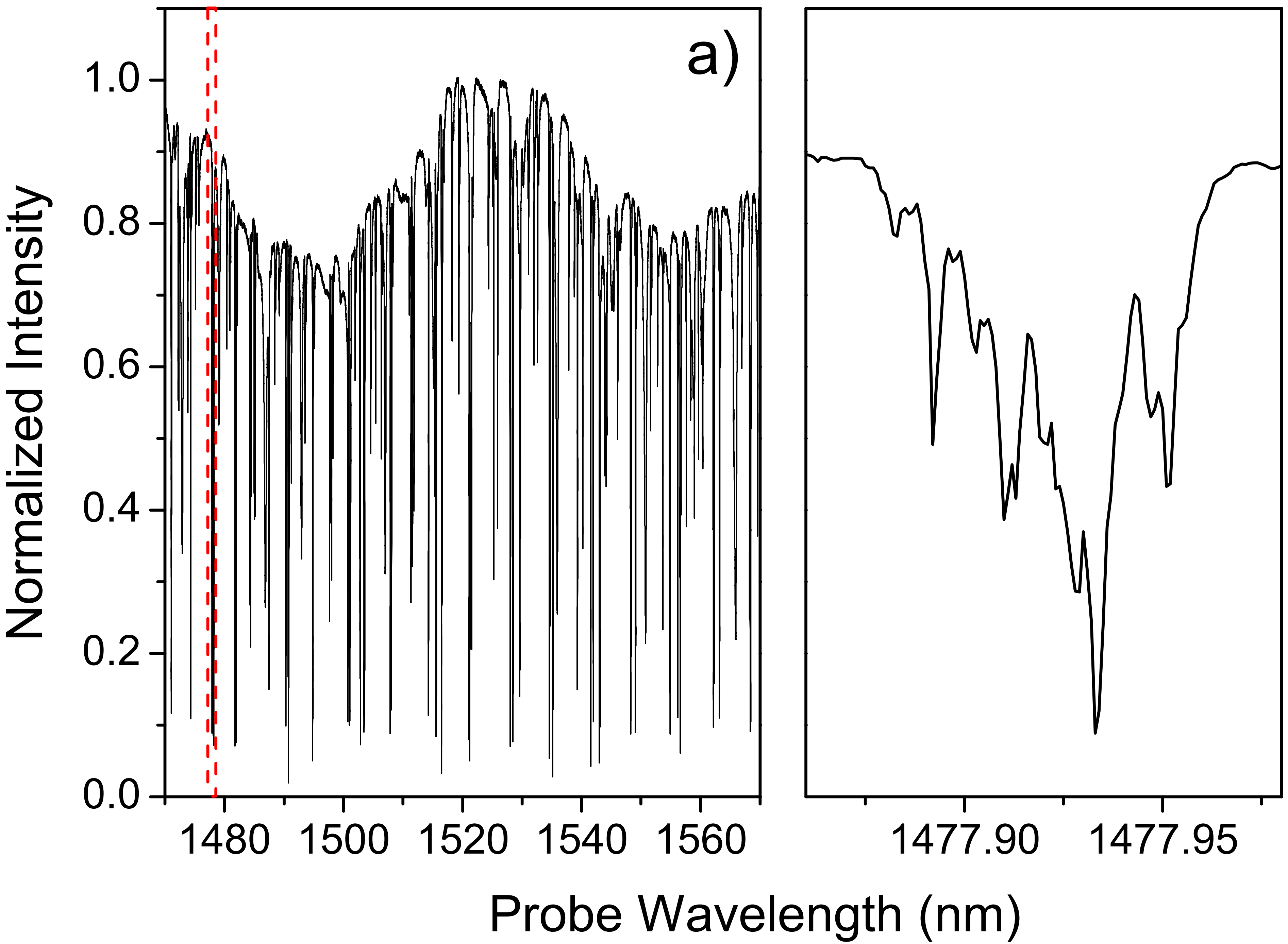

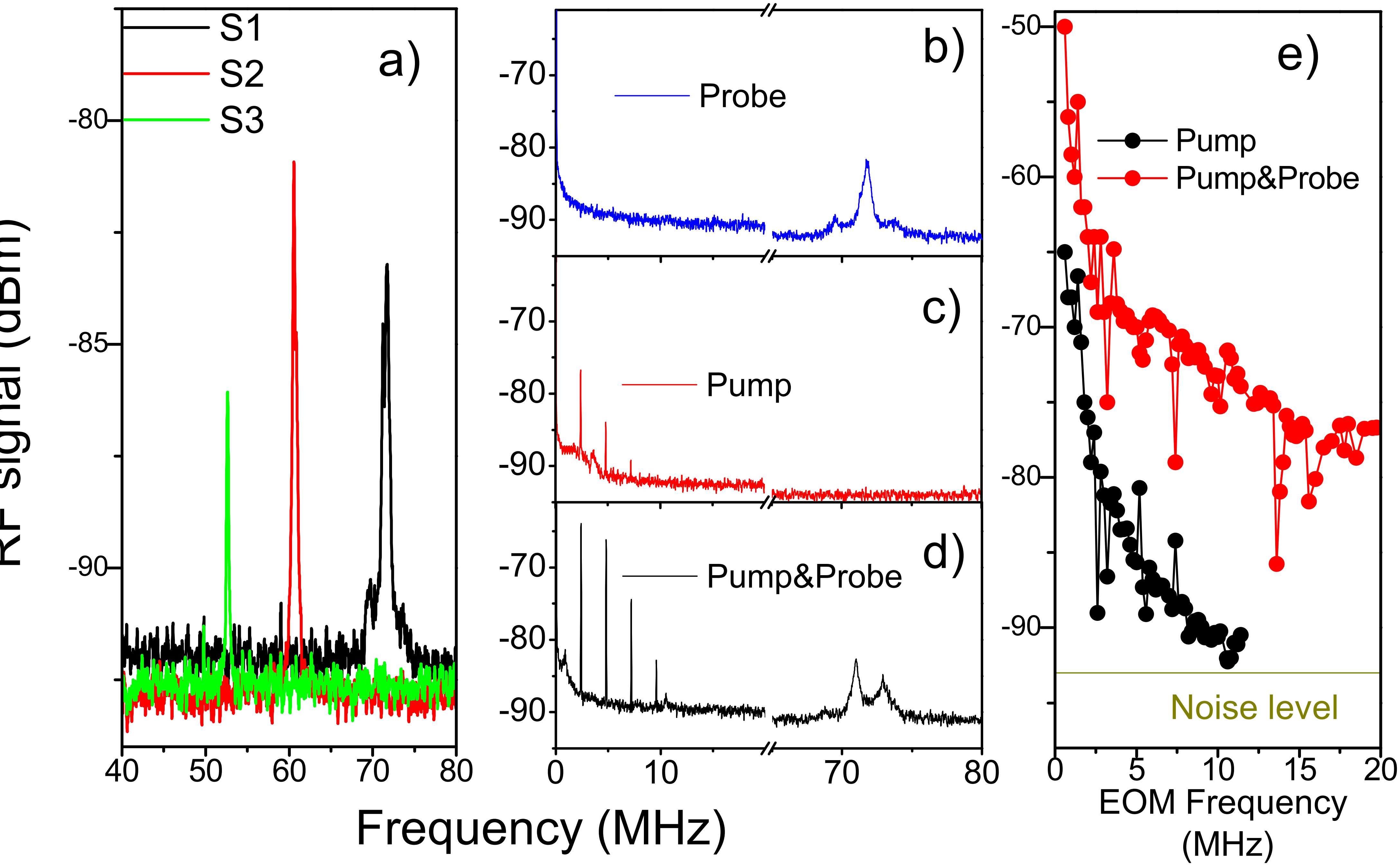

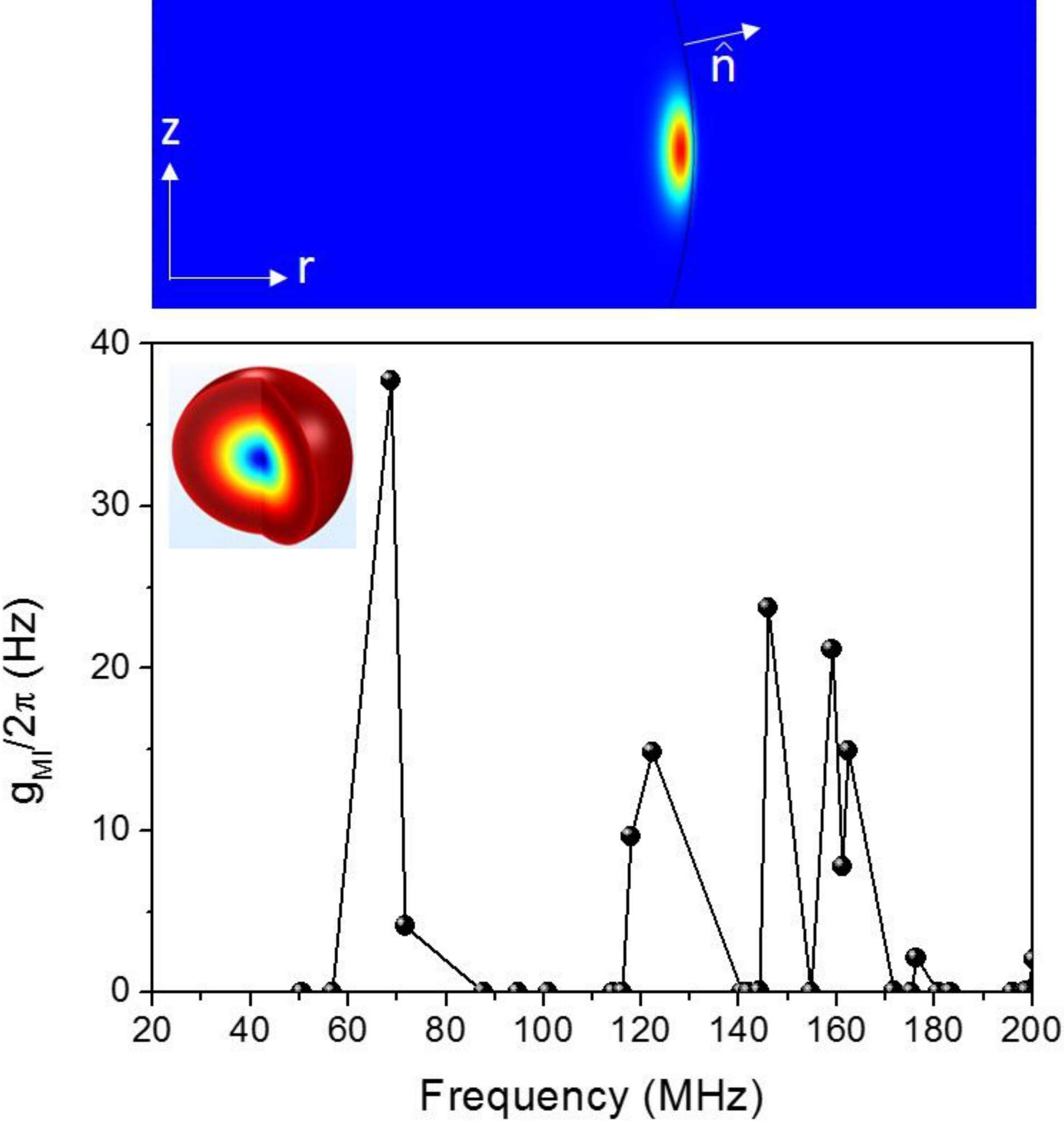